\documentclass[prl,twocolumn,showpacs]{revtex4}
\usepackage{graphicx}

\begin{document}
\tighten
%\documentstyle[12pt]{article}
%\documentclass[12pt]{article}
%\usepackage{graphics}
%\def\baselinestretch{1.5}
%\begin{document}
\newcommand {\be}{\begin{equation}}

\newcommand {\ee}{\end{equation}}

\newcommand {\bea}{\begin{eqnarray}}

\newcommand {\eea}{\end{eqnarray}}

\def\lsim{\:\raisebox{-0.5ex}{$\stackrel{\textstyle<}{\sim}$}\:}

\def\gsim{\:\raisebox{-0.5ex}{$\stackrel{\textstyle>}{\sim}$}\:}

\def\Dtens{\mbox{\sffamily\bfseries D}}

\def\Wtens{\mbox{\sffamily\bfseries W}}

\def\Ptens{\mbox{\sffamily\bfseries P}}

\def\Otens{\mbox{\sffamily\bfseries O}}

\def\Qtens{\mbox{\sffamily\bfseries Q}}

\def\Q{\mbox{\sffamily\bfseries Q}}

\def\Ntens{\mbox{\sffamily\bfseries N}}

\def\Ctens{\mbox{\sffamily\bfseries C}}

\def\Itens{\mbox{\sffamily\bfseries I}}

\def\Atens{\mbox{\sffamily\bfseries A}}

\def\A{\mbox{\sffamily\bfseries A}}

\def\Ktens{\mbox{\sffamily\bfseries K}}

\def\Vtens{\mbox{\sffamily\bfseries V}}

\def\Gtens{\mbox{\sffamily\bfseries G}}

\def\ftens{\mbox{\sffamily\bfseries f}}

\def\vtens{\mbox{\sffamily\bfseries v}}

\def\nabbold{\mbox{\boldmath $\nabla$\unboldmath}}

\def\nabvec{\mbox{\boldmath $\nabla$}}

\def\sigtens{\mbox{\boldmath $\sigma$\unboldmath}}

\def\etatens{\mbox{\boldmath $\eta$\unboldmath}}

\def\beq{\begin{equation}}

\def\bea{\begin{eqnarray}}

\def\eeq{\end{equation}}

\def\eea{\end{eqnarray}}

\title{The Fast Wandering of Slow Birds}

\author{John Toner}
\address{Department of Physics and Institute of Theoretical Science,
University of Oregon, Eugene, OR 97403, USA}

\date{\today}

\begin{abstract}

I study  a single ``slow'' bird moving with
a flock of birds of a different, and faster (or slower)  species. I
find that every ``species" of  flocker has a characteristic speed $\gamma\ne v_0$, 
where $v_0$ is the mean speed of the flock, such that, if  the
speed $v_s$ of the ``slow'' bird  equals $\gamma$, it will randomly wander 
transverse to
the mean direction
of flock motion far faster than the other birds will:  its mean-squared
transverse displacement will grow in $d=2$ with   time
$t$ like $t^\frac{5}{3}$, in contrast to $t^\frac{4}{3}$ for the other birds. In $d=3$, the slow bird's mean squared transverse displacement grows like $t^\frac{5}{4}$, in contrast to $t$ for the other birds. 
If $v_s\neq \gamma$, the mean-squared displacement of
the ``slow'' bird crosses over
from $t^\frac{5}{3}$ to
$t^\frac{4}{3}$ scaling in $d=2$, and from $t^\frac{5}{4}$ to
$t$ scaling in $d=3$, at a  time $t_c$ that scales according to
$t_c \propto|v_s-\gamma|^{-2}$. 
\end{abstract}


\maketitle

%Article-id: 0909.1954, Article password: pkjz5 (access no longer 
%password restricted)

One of the most familiar and ubiquitous biological phenomena,
flocking \cite{boids}   -- the collective, coherent motion of large numbers of
organisms --  also spans an enormous range of length scales:  from
kilometers (herds of wildebeest) to microns (e.g., the
microorganism Dictyostelium discoideum
\cite{dictyo,rappel1, rappel2}). Despite this familiarity, ubiquity, 
and enormous
range of length scales,  it is only in the last 15 years or so that 
the very rich
and exotic scaling phenomena that occur in flocks has begun to be appreciated.

This new understanding began with Vicsek's \cite{Vicsek} recognition that this
phenomenon is a dynamical version of ferromagnetic ordering. Building on this
insight,  a phenomenological continuum theory of flocking\cite{TT1,
TT2, TT3,TT4} was developed, which showed, among other things, that
two-dimensional flocks did not obey the Mermin-Wagner theorem
\cite{MW}: that is,
they {\it are} capable of spontaneously breaking a continuous
symmetry (rotation
invariance) by developing long-ranged order, even in flocks with only short
ranged interactions. Such order is impossible (in spatial dimensions $d=2$) for
equilibrium systems \cite{MW}.

This fundamental difference between flocks and  equilibrium systems arises from
novel ``anomalous'' scalings of fluctuations in flocks. Many of the
quantitative
predictions  of the phenomenological continuum theory about these scalings
have been confirmed in numerical simulations\cite{TT2,Chate1, Chate2}.

One probe of this anomalous scaling that seems particularly apt to anyone who's
ever watched a nature program in which  wild animals are  tagged with tracking
devices (and who among us has not?) is to study the wandering of one particular
creature flying along as part of a flock.The continuum theory\cite{TT1,
TT2, TT3,TT4} predicts that  the lateral wandering of a bird in a flock of
identical birds  grows superdiffusively in
$d=2$, scaling with time $t$  like $t^{\frac{2}{3}}$. The spectacular
confirmation of this very striking prediction in numerical
experiments\cite{TT2,Chate2} remains one of the strongest tests to date of the
analytic continuum theory.

However, such experiments only probe a small region of the full spatio-temporal
range of flock motion; specifically, the regime of spatio-temporal 
points moving
at the mean speed of the flock. It would clearly be a valuable, and more
rigorous, test of the theory to probe other regions as well.

In this paper, I suggest one way to do so: by including  a single
``slow'' bird, different from the other birds
in the flock, which attempts to follow the other birds, but flies  at
a speed
$v_s\neq v_0$, where $v_0$ is the average speed of the flock.  I find that each species of bird has a certain ``magic" speed $\gamma\neq v_0$ such that, if a ``slow" bird of a different species flying at a speed 
$v_s=\gamma$
is added to the flock, it will   wander
laterally much faster even than the superdiffusive wandering of the 
others. This
difference is a direct result of the different spatio-temporal 
regions probed by
this bird because it is not moving at the average speed of the flock.

%Most of this paper deals
%  with the two-dimensional case. At the end of the paper,  I will
%simply quote the results  for the three dimensional case. I will, 
%in addition, quote results for four and more dimensions, where
%a linearized
%version of the continuum theory is adequate\cite{log}.

My result applies to any ``ordered" flock moving on a substrate: i.e., one  in which the flocking organisms spontaneously
pick a direction
to move together via purely short-ranged interactions that make
neighbors tend to
follow each other, but which do {\it not} pick out any a priori
preferred direction
for this motion. That is, the flocking spontaneously breaks rotation
invariance, as
equilibrium ferromagnetism does. Flocks moving without a substrate conserve momentum, and so have a very different hydrodynamics, which has been considered elsewhere\cite{mom cons}. Whether or not slow birds wander rapidly in such a flock is an open question which remains uninvestigated, and which I will not address here.

One specific realization of such a system is one obeying the Vicsek
algorithm \cite{Vicsek}, in which birds attempt to follow their near
neighbors, but
make mistakes. This algorithm has a parameter range in which the
system ``flocks''
(i.e., has a non-zero average velocity:
$<\vec{v}>\neq \vec{0}$, where the average is over all the birds in a
very large flock
(i.e., in the thermodynamic limit)). In this ``ordered'' parameter
range, the continuum
theory \cite{TT1, TT2, TT3,TT4} applies.

I will now briefly review the salient features of this continuum theory.
This approach describes the flock by continuous, coarse grained 
number density $\rho(\vec{r}, t)$ and velocity $\vec{v}(\vec{r}, t)$ 
fields. The continuum equations of motion governing these fields
 can in the long-wavelength limit can be written down purely on 
 symmetry grounds, and are:\ \cite{TT1, TT2, TT3, TT4}
\begin{widetext}
\begin{eqnarray}
\partial_{t}
\vec{v}&+&\lambda_1 {(\rho,
|\vec{v}|)}(\vec{v}\cdot\vec{\nabla})\vec{v}+
\lambda_2 {(\rho, |\vec{v}|)}(\vec{\nabla}\cdot\vec{v})\vec{v}
+\lambda_3 {(\rho, |\vec{v}|)}\vec{\nabla}(|\vec{v}|^2)
=
\alpha{(\rho, |\vec{v}|)}\vec{v}-\beta{(\rho, |\vec{v}|)}
|\vec{v}|^{2}\vec{v} -\vec{\nabla} P{(\rho, |\vec{v}|)}  \nonumber \\&-&\vec{v} 
\left( \vec{v} \cdot \vec{\nabla}  P_2 {(\rho, |\vec{v}|)}\right) +
D_{B}{(\rho, |\vec{v}|)} \vec{\nabla}
(\vec{\nabla}
\cdot \vec{v})+ D_{T}{(\rho, |\vec{v}|)}\nabla^{2}\vec{v} +
D_{2}{(\rho, |\vec{v}|)}(\vec{v}\cdot\vec{\nabla})^{2}\vec{v}+\vec{f}~~,
\label{EOM}
\end{eqnarray}
\end{widetext}
\begin{eqnarray}
{\partial\rho \over \partial
t}+\nabla\cdot(\vec{v}\rho)=0~~,
\label{conservation}
\end{eqnarray}
where all of the parameters $\lambda_i (i = 1 \to 3)$,
$\alpha$, $\beta$, $D_{B,T,2}$ and the  ``isotropic Pressure'' $P(\rho,
|\vec{v}|)$ and the  ``anisotropic Pressure''$P_2 (\rho, |\vec{v}|)$
are, in general, functions of the density $\rho$ and the magnitude
$|\vec{v}|$ of the local velocity. It is useful to Taylor expand $P$ and 
$P_2$ around the equilibrium density $\rho_0$:
\begin{eqnarray}
P=P(\rho, |\vec{v}|)=\sum_{n=1}^{\infty} \sigma_n ( |\vec{v}|)
(\rho-\rho_0)^n ~~,
\label{P rho}
\end{eqnarray}
\begin{eqnarray}
P_2=P_2(\rho, |\vec{v}|)=\sum_{n=1}^{\infty} \mu_n ( |\vec{v}|)
(\rho-\rho_0)^n~~.
\label{P2 rho}
\end{eqnarray}

Here $\beta$, $D_{B}$, $D_{2}$ and $D_{T}$ are all
positive, and
$\alpha < 0$ in the disordered phase and $\alpha>0$ in
the ordered state (in mean field theory).  

The
$\alpha$ and
$\beta$ terms simply make the local
$\vec{v}$ have a nonzero magnitude $v_0=\sqrt{{\alpha} \over
{\beta}}$ \cite{implicit}  
in the ordered phase, where
$\alpha>0$. $D_{B,T,2}$ are the diffusion constants (or
viscosities) reflecting the tendency of a localized fluctuation in the
velocities to spread out because of the coupling between
neighboring ``birds".   The $\vec{f}$ term is a random
driving force representing the noise. It is  assumed to be Gaussian with
white noise correlations:
\begin{eqnarray}
   <f_{i}(\vec{r},t)f_{j}(\vec{r'},t')>=\Delta
\delta_{ij}\delta^{d}(\vec{r}-\vec{r'})\delta(t-t') ~~,
\label{white noise}
\end{eqnarray}
   where $\Delta$ is a constant, and $i$ , $j$ denote
Cartesian components. The pressure $P$ tends, as in an equilibrium fluid, 
to maintain the local number density
$\rho(\vec{r})$ at its mean value $\rho_0$,
and $\delta \rho = \rho -
\rho_0$.   The ``anisotropic pressure'' $P_2(\rho, |\vec{v}|)$ in
(\ref{EOM}) is only allowed due to the non-equilibrium nature of the
flock; in an equilibrium fluid such a term is forbidden, since Pascal's
Law ensures that pressure is isotropic. In the nonequilibrium steady
state of a flock, no such constraint applies. In earlier work\ \cite{TT1,
TT2, TT3, TT4} this term was ignored. While its inclusion does not affect
the linearized theory of the ordered flock, it could, potentially, have
important non-linear effects, as I discuss below.

The final
equation (\ref{conservation}) is just conservation of bird number: we
don't allow our birds to reproduce or   die on the wing.
The interesting and novel results that arise when this constraint
is relaxed by allowing birth and death while the flock is moving will be
discussed elsewhere\ \cite{Malt}.

The hydrodynamic model embodied in equations (\ref{EOM}),
(\ref{P rho}), and (\ref{conservation})  is equally valid  in both the
``disordered'' (i.e., non-moving) ($\alpha  < 0$) and 
``ferromagnetically ordered'' (i.e.,  moving) ($\alpha  > 0$) state . Here
I am interested in the ``ferromagnetically ordered'',
broken-symmetry phase which occurs for $\alpha>0$. In this state,   the velocity field can be written as:
\begin{eqnarray}
\vec{v}=v_{0}\hat{x}_{\parallel}+\vec{\delta v}= (v_{0}+\delta v_\parallel)\hat{x}_{\parallel}+\vec{ v}_\perp~~,
\label{v fluc}
\end{eqnarray}
 where
$v_{0}\hat{x}_{\parallel}=<\vec{v}>$ is the spontaneous average
value of
$\vec{v}$ in the ordered phase, and the fluctuations $\delta v_\parallel$ and $\vec{ v}_\perp$ of $\vec{v}$ about this mean velocity along and perpendicular to the direction of the mean velocity are assumed to be small. Indeed, I will be shortly be expanding the equation of motion (\ref{EOM}) in these quantities.
 Taking
$v_0=\sqrt{{\alpha} \over {\beta}}$ as discussed above \cite{implicit}, and taking the dot product of both sides of equation (\ref{EOM}) with $\vec{v}$ itself, I obtain:
%+\lambda_2 {(\rho, |\vec{v}|)}(\vec{\nabla}\cdot\vec{v})\vec{v}+\lambda_3 {(\rho, |\vec{v}|)}\vec{\nabla}(|\vec{v}|^2)=\alpha\vec{v}-\beta|\vec{v}|^{2}\vec{v} -\vec{\nabla} P -\vec{v} 
%\left( \vec{v} \cdot \vec{\nabla}  P_2 \right) +D_{B} \vec{\nabla}
%(\vec{\nabla}\cdot \vec{v}) \nonumber \\&+& D_{T}\nabla^{2}\vec{v} +D_{2}(\vec{v}\cdot\vec{\nabla})^{2}\vec{v}+\vec{f}
\begin{widetext}
\begin{eqnarray}
{1\over 2}\left(\partial_{t}|\vec{v}|^2+(\lambda_1 + 2 \lambda_3)(\vec{v}\cdot\vec{\nabla})|\vec{v}|^2\right) + \lambda_2(\vec{\nabla}\cdot\vec{v})|\vec{v}|^2&= &(\alpha-\beta|\vec{v}|^{2})|\vec{v}|^{2}-\vec{v} \cdot \vec{\nabla}  P-|\vec{v}|^{2}\vec{v} \cdot \vec{\nabla}  P_2 +D_{B} \vec{v}\cdot\vec{\nabla}
(\vec{\nabla}\cdot \vec{v}) \nonumber \\&+& D_{T}\vec{v}\cdot\nabla^{2}\vec{v} +D_{2}\vec{v}\cdot\left((\vec{v}\cdot\vec{\nabla})^{2}\vec{v}\right)+\vec{v}\cdot\vec{f}
\label{v parallel elim}~.
\end{eqnarray}
\end{widetext}
In this hydrodynamic approach, 
we are interested only in fluctuations $\vec{\delta v}(\vec{r}, t)$ and $\delta \rho(\vec{r}, t)$
that vary slowly in space and time. (Indeed, the hydrodynamic equations (\ref{EOM}) and (\ref{conservation}) are only valid in this limit). Hence, terms involving space and time derivatives of 
$\vec{\delta v}(\vec{r}, t)$ and $\delta \rho(\vec{r}, t)$
are always negligible, in the hydrodynamic limit, compared to terms involving the same number of powers of fields without any time or space derivatives.

Furthermore, the fluctuations 
$\vec{\delta v}(\vec{r}, t)$ and $\delta \rho(\vec{r}, t)$ can themselves be shown to be small in the long-wavelength limit. Hence, we need only keep terms in (\ref{v parallel elim}) up to linear order in 
$\vec{\delta v}(\vec{r}, t)$ and $\delta \rho(\vec{r}, t)$. The 
$\vec{v}\cdot\vec{f}$ term can likewise be dropped, since it only leads to a term of order 
$\vec{v}_\perp f_\parallel$ in the $\vec{v}_\perp$ equation of motion, which is negligible (since $\vec{v}_\perp$ is small) relative to the $\vec{f}_\perp$ term already there.

These observations can be used to eliminate many of the terms in equation (\ref{v parallel elim}), and solve for the quantity
\begin{eqnarray}
U \equiv ( \alpha(\rho , |\vec{v}|)-\beta (\rho ,
|\vec{v}|)|\vec{v}|^2) &;
\label{Udef}
\end{eqnarray}
%is readily seen to contain terms of  $O(\delta v_\parallel)$ (to leading order in $\vec{\delta v}$,  while $\partial_{t}|\vec{v}|^2, (\vec{v}\cdot\vec{\nabla})|\vec{v}|^2, $ and the $D$ terms are either $O(\partial_t\delta v_\parallel)$, $O($spatial derivatives of $~\delta v_\parallel)$, or involve higher powers of $\vec{v}_{\perp}$. Hence, I can drop all of these terms from (\ref{v parallel elim}).  A slightly more involved RG analysis shows that , as can terms involving higher than linear powers of $\vec{\delta v}$.
the  solution  is:
\begin{widetext}
\begin{eqnarray}
U=\lambda_2 \vec{\nabla}\cdot\vec{v}+\vec{v}\cdot\vec{\nabla}P_2+{\sigma_1\over v_0}\partial_\parallel \delta \rho+{1\over 2 v_0}\left(\partial_t + \gamma_2 \partial_\parallel \right)\delta v_\parallel ~~.
\label{Usol}
\end{eqnarray}
\end{widetext}
 Inserting this expression for $U$ back into equation (\ref{v parallel elim}) (where $U$ appears by virtue of its definition (\ref{Udef})),  I find that $P_2$ and $\lambda_2$ cancel out of the $\vec{v}$ equation of motion, leaving
\begin{widetext}
\begin{eqnarray}
\partial_{t}
\vec{v}+\lambda_1(\vec{v}\cdot\vec{\nabla})\vec{v}+\lambda_3 \vec{\nabla}(|\vec{v}|^2)
&=&{\sigma_1\over v_0}
\vec{v} (\partial_\parallel \delta \rho)-\vec{\nabla} P + D_{B} \vec{\nabla}
(\vec{\nabla}
\cdot \vec{v})+ D_{T}\nabla^{2}\vec{v} +
D_{2}(\vec{v}\cdot\vec{\nabla})^{2}\vec{v}\nonumber\\&+&\left[{1\over 2 v_0}\left(\partial_t + \gamma_2 \partial_\parallel \right)\delta v_\parallel
\right]\vec{v}+\vec{f}~~.
\label{EOM2}
\end{eqnarray}
\end{widetext}
This can be made into an equation of motion for $\vec{v}_\perp$ involving only $\vec{v}_\perp(\vec{r}, t)$ and $\delta \rho(\vec{r}, t)$ by projecting perpendicular to the direction of mean flock motion $\hat{x}_\parallel$, and eliminating $\delta v_\parallel$ using equation(\ref{Usol}) and 
the expansion 
\begin{eqnarray}
U\approx-\Gamma_1\left(\delta v_\parallel +{|\vec{v}_\perp|^2\over 2 v_0}\right) - \Gamma_2 \delta \rho ~~,
\label{Uexp}
\end{eqnarray}
where 
I've defined 
\begin{eqnarray}
\Gamma_1 \equiv -\left({\partial U
 \over \partial |\vec{v}|}\right)^0_{\rho} &,
&\Gamma_2 \equiv - \left({\partial U
 \over \partial \rho}\right)^0_{|\vec{v}|}~,
\label{gamma12 def}
\end{eqnarray}
with, here and hereafter , super-
or sub-scripts
$0$ denoting functions of  $\rho$ and 
$|\vec{v}|$ evaluated at $\rho = \rho_0$ and $ |\vec{v}|=v_0$.
I've also used the expansion
(\ref{v fluc}) for the velocity in terms of the fluctuations $\delta v_\parallel$ and $\vec{ v}_\perp$ to write
\begin{eqnarray}
|\vec{v}|=v_0+\delta v_\parallel +{|\vec{v}_\perp|^2\over 2 v_0}+O(\delta v_\parallel ^2, |\vec{v}_\perp|^4)~,
\label{speed}
\end{eqnarray}
and kept only terms that an RG analysis shows to be relevant in the long wavelength limit.
Inserting  (\ref{Uexp}) into (\ref{Usol})  gives:
\begin{widetext}
\begin{eqnarray}
-\Gamma_1\left(\delta v_\parallel +{|\vec{v}_\perp|^2\over 2 v_0}\right) - \Gamma_2 \delta \rho =\lambda_2 \vec{\nabla}_\perp\cdot\vec{v}_\perp+\lambda_2\partial_\parallel \delta v_\parallel+{(\mu_1 v_0^2+\sigma_1)\over v_0}\partial_\parallel \delta \rho+{1\over 2 v_0}\left(\partial_t + \gamma_2 \partial_\parallel \right)\delta v_\parallel~~,
\label{v par 1}
\end{eqnarray}
\end{widetext}
where I've kept only linear terms on the right hand side of this equation, since the non-linear terms are at least of order derivatives of $|\vec{v}_\perp|^2$, and hence negligible, in the hydrodynamic limit, relative to the  $|\vec{v}_\perp|^2$ term explicitly displayed on the left-hand side.

This equation can be solved iteratively for $\delta v_\parallel$ in terms of $\vec{v}_\perp$, $\delta \rho$, and its derivatives. To lowest (zeroth) order in derivatives, $\delta v_\parallel \approx -{\Gamma_1\over \Gamma_2} \delta\rho$. Inserting this into equation(\ref{v par 1}) everywhere it appears on the right hand side of that equation gives $\delta v_\parallel$ to first order in derivatives:
\begin{widetext}
\begin{eqnarray}
\delta v_\parallel\approx -{\Gamma_1\over \Gamma_2}\left( \delta\rho+{1\over2v_0\Gamma_1}\partial_t \delta \rho+{\lambda_4
\partial_\parallel \delta\rho\over\Gamma_1}\right)-{\lambda_2\over\Gamma_1} \vec{\nabla}_\perp\cdot\vec{v}_\perp-{|\vec{v}_\perp|^2\over 2 v_0} ~~.
\label{v par 2}
\end{eqnarray}
\end{widetext}

%prove to be irrelvant compared to projecting  define $\gamma_2$, etc.

%with
%\begin{eqnarray}
%U \equiv ( \alpha(\rho , |\vec{v}|)-\beta (\rho ,|\vec{v}|)|\vec{v}|^2)|\vec{v}| &,\label{Udef}\end{eqnarray}

%Note now that if we are interested in ``hydrodynamic'' modes, by
%which I mean modes for which frequency $\omega \rightarrow 0$
%as wave vector $q \rightarrow 0$, we can, in the hydrodynamic
%$(\omega, q \rightarrow 0)$ limit, neglect $\partial_t
%\delta v_{\parallel}$ relative to $\Gamma_1 \delta v_{\parallel}$ in(
%\ref{v parallel elim}).  The resultant equation can trivially be solved for $\delta v_{\parallel}$:
%\begin{eqnarray}\delta v_{\parallel} = -(\sigma_1/\Gamma_1) \partial_{\parallel} 
%\delta\rho -\frac{\Gamma_2}{\Gamma_1}\delta\rho  &.\label{delta parallel elim}\end{eqnarray}
 
Inserting (\ref{v fluc}), (\ref{speed}), and 
(\ref{v par 2}) into the equation of motion (\ref{EOM2}) for
$\vec{v}$, and projecting that equation perpendicular to the mean direction of flock motion $\hat{x}_\parallel$ and (\ref{conservation}) for 
$\delta\rho$,  gives, neglecting ``irrelevant'' terms:
\begin{widetext}
\begin{eqnarray}
\partial_{t} \vec{v}_{\perp} + \gamma\partial_{\parallel} 
\vec{v}_{\perp} &+& \lambda^0_1 \left(\vec{v}_{\perp} \cdot
\vec{\nabla}_{\perp}\right) \vec{v}_{\perp} =-g_1\delta\rho\partial_{\parallel} 
\vec{v}_{\perp}-g_2\vec{v}_{\perp}\partial_{\parallel}
\delta\rho-g_3\vec{v}_{\perp}\partial_t
\delta\rho -{c_0^2\over\rho_0}\vec{\nabla}_{\perp}
\delta\rho -\sigma_2\vec{\nabla}_{\perp}(\delta \rho^2)\nonumber\\&+&
D^0_{B\rm{eff}}\vec{\nabla}_\perp\left(\vec{\nabla}_\perp\cdot\vec{v}_\perp\right)+
D^0_T\nabla^{2}_{\perp}\vec{v}_{\perp} +
D^0_{\parallel}\partial^{2}_{\parallel}\vec{v}_{\perp}+\nu_t\partial_t\vec{\nabla}_{\perp}\delta\rho+\nu_\parallel\partial_\parallel\vec{\nabla}_{\perp}\delta\rho+\vec{f}_{\perp}  ~~,
\label{vEOMbroken}
\end{eqnarray}
\end{widetext}
where  
$D_{\rho} \equiv \rho_0{\sigma_1 \over \Gamma_1}$,
$D^0_{B\rm{eff}}\equiv D^0_b+{2v_0\lambda_3\lambda_4\over\Gamma_1}$, $D^0_T$, and $D^0_{\parallel} \equiv
D^0_{T}+D^0_{2}v_0^2$ are the diffusion constants, and I've defined
\begin{eqnarray}
\gamma \equiv
\lambda^0_1v_0~ , 
\label{gamma_1 def}
\end{eqnarray}
\begin{eqnarray}
g_1 \equiv\left(
{\partial\lambda_1
 \over \partial \rho}\right)_0 -{\Gamma_2\lambda_1\over\Gamma_1}~, 
 \label{g_1 def}
 \end{eqnarray}
 \begin{eqnarray}
g_2 \equiv {\sigma_1\over v_0} - 
{(\lambda_1+2\lambda_3)\Gamma_2\over 2\Gamma_1} ~~,
\label{g_2 def}
\end{eqnarray}
\begin{eqnarray}
g_3 \equiv {\Gamma_2\over 2v_0\Gamma_1} ~~,
\label{g_3 def}
\end{eqnarray}
\begin{eqnarray}
c_0^2 \equiv \rho_0\sigma_1 - {2\rho_0v_0\lambda_3\Gamma_2\over \Gamma_1} ~~,
\label{c_0^2 def}
\end{eqnarray}
\begin{eqnarray}
\nu_t \equiv {\Gamma_2\lambda_3\over \Gamma_1^2} ~~,
\label{nu_t def}
\end{eqnarray}
\begin{eqnarray}
\nu_\parallel\equiv {2\Gamma_2v_0\lambda_3\lambda_4\over \Gamma_1^2} ~~,
\label{nu par def}
\end{eqnarray}
\begin{eqnarray}
D_{B\rm{eff}} \equiv D_B^0 + {2v_0\lambda_2\lambda_3\over \Gamma_1} ~~,
\label{D_{B} def}
\end{eqnarray}
and, finally, 
\begin{eqnarray}
D_\parallel \equiv D_T + D_2v_0^2 ~~.
\label{D_|| def}
\end{eqnarray}

It is extremely important to note that the speed $\gamma$ is {\it not} equal to the mean speed of the flock, since 
$\lambda_1\ne 1$. The reason that $\lambda_1\ne 1$ is the lack of Galilean invariance in the system under consideration, which in turn is due to the fact that we are considering flocks that move over a fixed surface, which provides a special rest frame. 

This symmetry argument is confirmed by more microscopic, kinetic theory approaches\cite{kin theory}, which indeed find $\lambda\ne 1$. Furthermore, simulations can determine $\lambda_1$ by fitting the predictions of this hydrodynamic model to the observed correlation functions. The fits work extremely well, and find $\lambda_1\ne 1$ \cite{TT2}.

So $\gamma=\lambda_1 v_0\ne v_0$;  as we will see in a moment, this means that the identical birds making up a flock, which of course all move at speed $v_0$, will wander laterally (i.e., perpendicular to the direction of mean motion $\hat{x}_\parallel$) much more slowly than a ``slow" bird whose speed $v_s=\gamma$. Note that since no general argument requires $\lambda_1 < 1$, the ``slow" bird might actually have to move {\it faster} than the ordinary birds, for some species.
 
Using  (\ref{v fluc}) and (\ref{speed}) in the equation of motion (\ref{conservation}) for $\rho$ gives, again neglecting irrelevant terms:
\begin{widetext}
\begin{eqnarray}
\partial_t\delta
\rho +\rho_o\vec{\nabla}_\perp\cdot\vec{v}_\perp
+\vec{\nabla}_\perp\cdot(\vec{v}_\perp\delta\rho)+v_2
\partial_{\parallel}\delta
\rho =D_{\rho\parallel}\partial^2_\parallel\delta\rho+D_{\rho v} \partial_{\parallel}\left(\vec{\nabla}_\perp \cdot \vec{v}_{\perp}\right)+w_1\partial_t\partial_\parallel\delta\rho+w_2\partial_\parallel(\delta\rho^2) ~~,
\label{cons broken}
\end{eqnarray}
\end{widetext}
where I've defined:

\begin{eqnarray}
v_2 \equiv v_0 - {\rho_0\Gamma_2\over \Gamma_1} ~~,
\label{v_2 def1}
\end{eqnarray}
\begin{eqnarray}
w_1 \equiv {\Gamma_2\rho_0\over 2v_0\Gamma_1^2} ~~,
\label{w_1 def}
\end{eqnarray}
\begin{eqnarray}
D_{\rho\parallel}\equiv {\Gamma_2\over \Gamma_1^2} \left({\lambda_1\over 2} + \lambda_2 + \lambda_3 \right) \rho_0 ~~,
\end{eqnarray}
\begin{eqnarray}
D_{\rho v} \equiv {\lambda_2\rho_o\over \Gamma_1} ~~,
\label{D_rho v def}
\end{eqnarray}
and, last but not least, 
\begin{eqnarray}
w_2 \equiv {\Gamma_2\over 2\Gamma_1} ~~.
\label{w_2 def}
\end{eqnarray}

%Note that $v_2\ne v_0$, unless $\Gamma_2=0$, which it will not, in general, since there is no symmetry that requires $\Gamma_2$ to vanish. It is because $v_2\ne v_0$ that a ``slow" bird moving at this speed $v_2$ wanders more rapidly than the other birds in the flock,  as we'll see below.
 
%The pressure $P$ continues to be given, as it always will, by equation (\ref{P rho}).

%For the purposes of obtaining the asymptotic long-wavelength,  long-time behavior of the ordered state of the flock, it is sufficient to replace each of the $\sigma_n$'s  and $\mu_n$'s in those equations with $\sigma_n^0$  and $\mu_n^0$ which are constants. Furthermore, higher order terms in the expansions of the various functions of 
%$\rho$ and $ |\vec{v}| $ appearing in equations (\ref{EOM}), (\ref{P rho}), (\ref{P2 rho}) ,  and 
%(\ref{conservation}), prove to be irrelevant to the
%long-wavelength, long-time  behavior of the ordered state of the flock.

Linearizing (\ref{vEOMbroken}) and (\ref{cons broken}), and spatio-temporally Fourier transforming them makes it straightforward algebra to calculate the hydrodynamic mode structure and spatio-temporally Fourier transformed correlation functions of the flock. (To obtain the latter simply involves  using the assumed spatio-temporal noise correlations (\ref{white noise})). 

The mode structure consists of a pair of propagating sound modes, with direction-dependent speeds given by:
$c_{\pm}\left(\theta_{\vec{q}}\right)$ are given by
\begin{eqnarray}
c_{\pm}\left(\theta_{\vec{q}} \right) =
\left({\gamma + v_2 \over 2}\right)\cos
\left(\theta_{\vec{q}}
\right)
\pm c_2\left(\theta_{\vec{q}} \right) &,
\label{cplusminus}
\end{eqnarray}
where I've defined
\begin{eqnarray}
c_2\left(\theta_{\vec{q}} \right) \equiv
\sqrt{{\left(\gamma -v_2\right)^2 \cos^2
\left(\theta_{\vec{q}}
\right) \over 4} + c_0^2
\sin^2 \left(\theta_{\vec{q}} \right)} \quad ,
\label{c2}
\end{eqnarray}
where $\theta_{\vec{q}}$ is the angle between
$\vec{q}$ and the direction of flock motion (i.\ e., the
$x_\parallel$ axis).

I'm particularly interested in the velocity auto-correlation, since, as I'll show in a moment, this determines the wandering of the birds relative to the mean motion of the flock. In the linearized approach just described, I find the spatio-temporally Fourier transformed velocity-velocity autocorrelation function is, ignoring negligible terms: 
\begin{widetext}
\begin{eqnarray}
C_{\perp}
\left(\vec{q},\omega\right) \equiv\left<\vec{v}\left(\vec{q},\omega\right)\cdot\vec{v}\left(-\vec{q},-\omega\right)\right>&=&
{\Delta
\left(\omega - v_2q_{\parallel}\right)^2
\over
\left[(\omega-c_+\left(\theta_{\vec{q}}\right)q)^2+\epsilon_+^2(\vec{q})\right] 
\left[(\omega-c_-\left(\theta_{\vec{q}}\right)q)^2+\epsilon_-^2(\vec{q})\right] }\nonumber\\&+&{(d-2)\Delta
\over
\left[(\omega-\gamma q_\parallel)^2+\Gamma_T^2(\vec{q})\right] }
\label{CL}
\end{eqnarray}
\end{widetext}
where I've defined
%\begin{widetext}
\begin{eqnarray}
\epsilon_{\pm}\equiv{\rm{NUM}\over(2c_{\pm}(\theta_{\vec{q}})-(v_2+\gamma)\cos(\theta_{\vec{q}}))}
\label{damping 1}
\end{eqnarray}
with
\begin{widetext}
\begin{eqnarray}
\rm{NUM}&\equiv&(\Gamma_L(\vec{q}) +
\Gamma_\rho(\vec{q})
-w_1 c_{\pm}(\theta_{\vec{q}})\cos(\theta_{\vec{q}})q^2) c_{\pm}(\theta_{\vec{q}})-v_2\Gamma_L(\vec{q})\cos(\theta_{\vec{q}})-\gamma(\Gamma_\rho(\vec{q})-w_1 c_{\pm}(\theta_{\vec{q}})\cos(\theta_{\vec{q}})q^2)\cos(\theta_{\vec{q}})\nonumber\\
&+&{c_0^2\over \rho_0}D_{\rho v}{q_\parallel q_\perp^2\over q}-\rho_0 q_\perp^2(\nu_t c_{\pm}(\theta_{\vec{q}})+\nu_\parallel \cos(\theta_{\vec{q}}))
%q_{\parallel}\left(v_2 \Gamma_L\left(\vec{q}\right) +\gamma\Gamma_{\rho}
%\left(\vec{q}\right) \right)\right)^2
\label{damping}
\end{eqnarray}
\end{widetext}
and the
wavevector dependent  longitudinal, transverse, and
$\rho$ dampings
$\Gamma_{L,\rho}$ are given by:
\begin{eqnarray}
\Gamma_L\left(\vec{q}\right)  = D_L q_\perp^2 + D^0_{\parallel}
q^2_{\parallel} \, ,
\label{GLdef}
\end{eqnarray}
\begin{eqnarray}
\Gamma_T\left(\vec{q}\right)  = D_T^0 q_\perp^2 + D^0_{\parallel}
q^2_{\parallel} \, ,
\label{GTdef}
\end{eqnarray}
and
\begin{eqnarray}
\Gamma_{\rho}\left(\vec{q}\right) = D_{\rho\parallel} q^2_{\parallel} \, ,
\label{Grhodef}
\end{eqnarray}
with $D_L\equiv D^0_{B\rm{eff}}+D_T^0$.

The first term in equation 35 comes from the ``longitudinal" component $v_L$ of $\vec{v}_\perp$, by which I mean the component along $\vec{q}_\perp$, while the second comes from the $d-2$ ``transverse" components of $\vec{v}_\perp$ which are 
perpendicular to $\vec{q}_\perp$. Clearly, in $d=2$, only the longitudinal component is present; this is why the second (transverse) term in (\ref{CL}) vanishes in $d=2$.

%NL'S$\rightarrow$ DOM'D BY $Q_PAR~Q_PERP^ZETA$ AND $OMEGA=GAMMA QPAR=>\rightarrow$ SIMPLIFIES TO
What about the effect of the non-linear terms in (\ref{vEOMbroken}) and (\ref{cons broken})?   In spatial
dimensions $d<4$, these terms are
strongly 
relevant in the RG sense \cite{TT1} - \cite{TT4}.

What ``relevant in the RG sense'' means in plain English is that these
non-linear terms lead to  different scaling behavior at long
distances and times than predicted by the linearized version of
(\ref{vEOMbroken}) and (\ref{cons broken}), which of course ignores those
terms. 

Determining 	quantitatively this modified scaling 
is difficult; indeed, the only known analytic approach is an
$\epsilon \equiv 4 - d$ expansion about the critical dimension $d = 4$, which clearly tells us little about
what goes on in $d = 2$.

However, there are many examples known in which non-linearities that become relevant for $d$ just below $4$ become irrelevant again as $d$ is decreased  to $3$, and remain so down to $d=2$  (e.g., cubic symmetry breaking terms at  the phase transition of the $XY$ 
(O(2)) model\cite{XY cubic}). 

Now, in our problem, it is impossible for {\it all} of the non-linearities in (\ref{vEOMbroken}) and (\ref{cons broken}) to be irrelevant in any $d<4$, since the linearized fixed point can be shown to be unstable against {\it all} of these nonlinearities for those $d$'s. However, it is quite possible, as illustrated by the discussion of the previous paragraph, that the ordered phase of the flock is controlled by a fixed point in which {\it only one} of these non-linearities is important. While there is no analytic calculation so far that shows this is the case, if we {\it conjecture} that the dominant 
non-linearity in equation\ (\ref{vEOMbroken}) is $\lambda ^0_1$, and that {\it
all} of the other non-linearities are {\it irrelevant}, in the RG sense,
below
$d = 4$ (a result which is, at least, not ruled out by the $4 - \epsilon$
calculation that {\it has} been done), then one can show
\cite{TT1}-\cite{TT4} that the effect of these non-linearities on all spatio-temporally Fourier transformed two point correlation functions can be incorporated entirely by replacing $D^0_T$ and $ D_{B\rm{eff}}$ everywhere they appear in the linearized expressions for those correlation functions with renormalized, strongly wavevector ($\vec{q}$)-dependent
diffusion coefficients $D_{T,B}(\vec{q})$ that diverge as $\vec{q}\rightarrow\vec{0}$ according to the scaling law:
\begin{eqnarray}
&D_{T,B}({\vec q}) = q^{2\zeta-2}_{\perp} f_{T,B}
\left({\left({q_{\parallel}\over \Lambda}\right) \over
\left({q_{\perp} \over \Lambda} \right)^{\zeta} }\right)\nonumber
\\&\propto
\left\{\begin{array}{ll}
\mbox{q}_\perp^{2\zeta-2},& {q_\parallel\over\Lambda}\ll
({q_\perp\over\Lambda})^\zeta \\ q_\parallel^{ 2-{2\over\zeta}
},&{q_\parallel\over\Lambda}\gg ({q_\perp\over\Lambda})^\zeta
\end{array}\right.
\label{Dcross} ,
\end{eqnarray}
here $\Lambda$  is an ultraviolet cutoff of order an inverse inter-bird
distance, and the universal exponent $\zeta$ is given by:
%exponents $z$, $\zeta$, and $\chi$ are given {\it exactly} , in all dimensions $2 \leq d \leq 4$, byREWRITE THE FOLLOWING AS $D_T(Q), D_B(Q)$ scaling like 
%$q_\perp^{\zeta-2}$, AND ALSO MENTION THAT $D_{\rho\perp}$
%IS GENERATED
%\begin{eqnarray}
%z = 2{(d + 1) \over 5}\label{zcan}\end{eqnarray}
\begin{eqnarray}
\zeta =  {d + 1 \over 5}
\label{zetcan}
\end{eqnarray}
In addition, $\Gamma_\rho(\vec{q})$ is also modified, becoming
\begin{eqnarray}
\Gamma_{\rho}\left(\vec{q}\right) = D_{\rho\parallel} q^2_{\parallel} +D_{\rho\perp}\left(\vec{q}\right) q^2_{\perp}\, ,
\label{Grhodef2}
\end{eqnarray}
where the new diffusion coefficient $D_{\rho\perp}\left(\vec{q}\right)$ also becomes 
anomalous (i.e., a divergent function of wavevector $\vec{q}$ as $\vec{q}\rightarrow
\vec{0}$; indeed, it obeys the   same scaling law 
as $D_{T,B}$; i.e.,
\begin{eqnarray}
&D_{\rho\perp}({\vec q}) = q^{2\zeta-2}_{\perp} f_{\rho\perp}
\left({\left({q_{\parallel}\over \Lambda}\right) \over
\left({q_{\perp} \over \Lambda} \right)^{\zeta} }\right)\nonumber
\\&\propto
\left\{\begin{array}{ll}
\mbox{q}_\perp^{2\zeta-2},& {q_\parallel\over\Lambda}\ll
({q_\perp\over\Lambda})^\zeta \\ q_\parallel^{ 2-{2\over\zeta}
},&{q_\parallel\over\Lambda}\gg ({q_\perp\over\Lambda})^\zeta
\end{array}\right.
\label{Dcross2} .
\end{eqnarray}

As a result of the divergences of $D_{B,T,\rho\perp}({\vec q}) $, the dampings $\Gamma_{L,T,\rho}({\vec q}) $ also obey scaling laws:
\begin{eqnarray}
&\Gamma_{L,T,\rho}({\vec q})  = q^{2\zeta}_{\perp} f_{\Gamma_L, \Gamma_T, \Gamma_\rho}
\left({\left({q_{\parallel}\over \Lambda}\right) \over
\left({q_{\perp} \over \Lambda} \right)^{\zeta} }\right)\nonumber
\\&\propto
\left\{\begin{array}{ll}
\mbox{q}_\perp^{2\zeta},& {q_\parallel\over\Lambda}\ll
({q_\perp\over\Lambda})^\zeta \\ q_\parallel^{2{\over}},
&{q_\parallel\over\Lambda}\gg ({q_\perp\over\Lambda})^\zeta
\end{array}\right.
\label{Gamma scale} .
\end{eqnarray}

%\begin{eqnarray}\chi =  {3 -2d  \over 5} \quad .\label{chican}\end{eqnarray}

While the reader may with some justice be skeptical that we would be so lucky as to have the only relevant non-linearity in the problem prove to be the only one that we can handle  analytically, there is considerable evidence from simulations\cite{TT2,TT3,Chate1, Chate2} supporting the ``canonical" behavior just described, in both $d=2$ and $d=3$. I will therefore assume for the remainder of this paper that this ``canonical" behavior is the true behavior of a flock.

The divergence as  $\vec{q}\rightarrow\vec{0}$ of the diffusion coefficients $D_{B,T,\rho\perp}(\vec{q})$ embodied in equations (\ref{Dcross}) and (\ref{Dcross2})  has dramatic consequences for the velocity fluctuations. In particular, it implies that, for fixed $|\vec{q}|$, these fluctuations will damp out far more slowly for $\vec{q}$ along 
the direction of mean flock motion $\hat{x}_\parallel$ than for any other direction of $\vec{q}$. More precisely, those modes with ${q_\parallel\over\Lambda}\sim ({q_\perp\over\Lambda})^\zeta$ will decay far more slowly than
those with ${q_\parallel\over\Lambda}\ll ({q_\perp\over\Lambda})^\zeta$.  This 
means that, as $\vec{q}\rightarrow\vec{0}$, the largest fluctuations will occur at 
wavevectors $\vec{q}$ with $q_\parallel\gg q_\perp$, since $\zeta<1$. Thus, we 
only need to evaluate $C_\perp(\vec{q}, \omega)$ for those values of $\vec{q}$ 
with  $q_\parallel\gg q_\perp$. In this limit, the
autocorrelation function (\ref{CL}) can be considerably simplified, to:
\begin{widetext}
\begin{eqnarray}
C_{\perp}
\left(\vec{q},\omega\right)\approx
{\Delta
\over
\left[(\omega-\gamma q_\parallel-{c_0^2 q_\perp^2\over (\gamma-v_2) q_
\parallel})^2
+\Gamma_L^2(\vec{q})\right]}+{(d-2)\Delta
\over
\left[(\omega-\gamma q_\parallel)^2+\Gamma_T^2(\vec{q})\right] }~.
\label{CL2}
\end{eqnarray}
\end{widetext}

Further simplification can be made by comparing the $c_0^2 q_\perp^2$ term in the first denominator with the $\Gamma_L^2(\vec{q})$ term therein. The ratio of these scales like
\begin{eqnarray}
{q_\perp^4\over q_
\parallel^2
\Gamma_L^2(\vec{q})}
\propto {q_\perp^{2(1-\zeta)}\over q_\parallel^2f_{\Gamma_L}{\left(\left({q_{\parallel}
\over \Lambda}\right) \over
\left({q_{\perp} \over \Lambda} \right)^{\zeta} \right)}}
\label{ratio1}
\end{eqnarray}

For the important regime of wavevectors $q_\perp^\zeta\sim q_\parallel$, this ratio scales\cite{x sim 1} like $q_\perp^{2-3\zeta}$, and, hence, vanishes at small $q_\perp$ if and only if the exponent $\zeta<{2\over 3}$. From equation (\ref{zetcan}), it's clear that this condition on $\zeta$  is satisfied in $d=2$, but not $d=3$. Thus, in $d=2$, where the second term in  (\ref{CL}) vanishes, we can drop the $c_0^2 q_\perp^2$ term in  (\ref{CL2}) , and obtain the much simpler expression:
\begin{eqnarray}
C_{\perp}
\left(\vec{q},\omega\right)\approx
{\Delta
\over
\left[(\omega-\gamma q_\parallel)^2
+\Gamma_L^2(\vec{q})\right]}~~~~~~~(d=2).
\label{CLd=2}
\end{eqnarray}

In $d=3$, on the other hand, the $c_0^2 q_\perp^2$ term in  (\ref{CL2})  dominates the $\Gamma_L(\vec{q})$ term, since $\zeta={4\over 5}$ in $d=3$. In this case, the first (i.e., the longitudinal) term in  (\ref{CL2})  is, throughout most of the important range of $\vec{q}$,  much smaller than it would be in the absence of this term. On the other hand, the second, $d-2$ term in  (\ref{CL2})  is comparable in magnitude to what the first term {\it would} have been, {\it had } I dropped the $c_0^2 q_\perp^2$ term. To see this, simply note that the two terms would then become identical, except for the factor of $d-2$, which is $1$ in $d=3$,  and the replacement of $\Gamma_L$ by $\Gamma_T$. Since $\Gamma_L$ and $\Gamma_T$ both scale the same way with $\vec{q}$, this latter change makes little difference. Hence, keeping the $c_0^2 q_\perp^2$ term makes the first (longitudinal) term in  (\ref{CL2})  actually much smaller than the second, transverse, one. Hence we can drop that first term, and write:
\begin{eqnarray}
C_{\perp}
\left(\vec{q},\omega\right)\approx
{\Delta
\over
\left[(\omega-\gamma q_\parallel)^2
+\Gamma_T^2(\vec{q})\right]}~~~~~~~(d=3).
\label{CTd=3}
\end{eqnarray}

Now let us add to this ordered  flock a single ``slow''
bird\cite{slow} which tries to
follow the other birds in the flock via some short-ranged
interaction, but which moves
at a different speed $v_s$ than the other birds in the flock. For
example, one could
imagine (or  simulate) a Vicsek algorithm\cite{Vicsek} in which the
``slow'' bird obeys the same algorithm as the others, except that its speed of
motion between time steps  differs from those of the other birds.

Consider now the mean squared lateral displacement of the slow bird:
\begin{eqnarray}
w^2(t; v_s) \equiv \left<\left|\vec{x}^{\perp}_s (t) -
\vec{x}^{\perp}_s (0) \right|^2
\right>
\label{RW 1}
\end{eqnarray}
{\it perpendicular} to the mean direction of motion of the flock. (Here
and throughout this paper, $\perp$ and $\parallel$ denote directions
perpendicular
to, and  along, the direction of mean flock motion, respectively.)

Since the mean
$\vec{x}_{\perp} -
$ position $\vec{x}^{\perp}_s$ of the slow bird obeys
\begin{eqnarray}
\vec{x}^{\perp}_s (t) =  \vec{x}^{\perp}_s (0) + \int^t_0
\vec{v}^{\perp}_s (t) dt
\label{RW 2}
\end{eqnarray}
where $v^{\perp}_s (t)$ is the ${\perp}$ velocity of the slow
bird  at time $t$, $w^2 (t; v_s)$ is given by
\begin{eqnarray}
w^2 (t; v_s) = \int^t_0 d t^{\prime} \int^t_0 dt^{\prime\prime}
\left< \vec{v}^{\perp}_s (t^{\prime}) \cdot \vec{v}^{\perp}_s
(t^{\prime\prime})\right> \quad .
\label{RW 3}
\end{eqnarray}
Now I need to relate the perpendicular velocity of the slow bird to the
position and time dependent velocity field $\vec{v}_{\perp}\left(\vec{r}, t
\right)$ of the flock.
Since the
slow bird moves in the same direction\cite{small_correction} as the other
birds in the flock, the required perpendicular velocity is given by:
\begin{eqnarray}
\vec{v}^{\perp}_s (t) = \frac{\vec{v}_{\perp} \left( \vec{r}_s (t), t\right)
v_s } {v_0}
\label{RW 4}
\end{eqnarray}
where $v_0$ is the
mean speed of the flock (in the sense that, $<\vec{v}(\vec{r}, t)> = v_0
\hat{x}_\parallel$, with $\hat{x}_\parallel$ the unit vector along
the direction of
mean flock motion), and
$\vec{r}_s (t)$ is the position of the slow bird at time
$t$. This is given by
\begin{eqnarray}
\vec{r}_s (t) = \vec{r}_s(0) + v_st \hat{x}_{\parallel} +
\delta x^{\parallel}_s(t) \hat{x}_{\parallel} + \vec{\delta{ x}}^{\perp}_s
(t)  \quad  ,
\label{RW 5}
\end{eqnarray}
where $\delta x^{\parallel}_s(t) $ and $ \vec{\delta{ x}}^{\perp}_s(t)$ are the
deviations of  the slow bird from uniform motion at  speed
$v_s$ along $\hat{x}_{\parallel}$.

   Using (\ref{RW 4}) and (\ref{RW 5}), I see that the desired single
bird autocorrelation function in (\ref{RW 3}) is:
%\newpage
\begin{widetext}
\begin{eqnarray}
\left< \vec{v}^{\perp}_s \left(t^{\prime}\right) \cdot  \vec{v}^{\perp}_s
\left(t^{\prime\prime}\right)\right> &=&
\left(\frac{v_s}{v_0}\right)^2 \left< \vec{v}_{\perp} \left(\vec{r}_s(t^\prime),t\right)\cdot
\vec{v}_{\perp} \left(\vec{r}_s(t^{\prime\prime}), t^{\prime\prime}) \right)\right> \nonumber\\
&=& \left(\frac{v_s}{v_0}\right)^2C_{\perp} \left(\vec{\delta x}_s^{\perp} 
\left(t^{\prime}\right)
-\vec{\delta x}_s^{\perp}\left(t^{\prime\prime}\right), v_s( t^{\prime} - t^{\prime\prime} ) +\delta x_s^{\parallel}\left(t^{\prime}\right) - \delta
x_s^{\parallel}\left(t^{\prime\prime}\right), t^{\prime} -
t^{\prime\prime}\right)
\label{RW 6}
\end{eqnarray}
\end{widetext}
%\newpage
where $C_\perp\left(\vec{r}_\perp, r_\parallel, t\right)$ is the real space and time perpendicular
velocity field auto-correlation function, and I've used
\begin{eqnarray}
\vec{r}_s(t)=\vec{r}_s(0) +
\left(v_st +
\delta x_s^{\parallel}\left(t\right)\right)\hat{x}_{\parallel}
+
\vec{\delta x}_s^{\perp}\left(t\right)~~,
\end{eqnarray}
to write both $\vec{r}_s(t^\prime)$ and $\vec{r}_s(t^{\prime\prime})$.

I'll assume (and have  verified a posteriori) that both $\delta x_s^{\parallel}$
and
$\delta x_s^{\perp}$ are small enough compared to the average
motion $v_st \hat{x}_{\parallel}$ that their effect on the
velocity-velocity autocorrelation in (\ref{RW 6}) is negligible. This will be true if $w^2(t)\ll t^2$, which, as I will show a posteriori, it is.
This leaves the task of evaluating
$C(\delta t)\equiv C_\perp \left(R_{\perp} =
0, R_{\parallel} = v_s\delta t, \delta t
\right)$, where I've defined $\delta t\equiv t^\prime - t^{\prime\prime}$.
Expressing $C_\perp$ in terms of its spatio-temporal Fourier transform  equation  (\ref{CL})  gives
\begin{widetext}
\begin{eqnarray}
C(\delta t)\equiv\left< \vec{v}^{\perp}_s \left(t^{\prime}\right)\cdot   \vec{v}^{\perp}_s
\left(t^{\prime\prime}\right)\right>=\left(\frac{v_s}{v_0}\right)^2C_\perp
\left(R_{\perp} = 0,    R_{\parallel} = v_s \delta t\right) =
\left(\frac{v_s}{v_0}\right)^2\int \frac{d^{d-1} q_{\perp} dq_{\parallel}
d\omega}{(2\pi)^{d+1}} \, e^{i\left(\omega -
v_s q_{\parallel} \delta t
\right)} C_\perp \left(\vec{q}, \omega \right) ~~ .
\label{Cv 1}
\end{eqnarray}
\end{widetext}
%where I've defined $\delta t \equiv t^\prime-t^{\prime\prime}$, and  the
%spatio-temporally perpendicular Fourier-transformedvelocity-velocity correlation
%function $C_\perp\left(\vec{q},
%\omega \right)$ is predicted by the continuum theory \cite{TT3,v2} to be:
%where $\Delta$ is a phenomenological parameter related to the RMS
%magnitude of thedirectional errors the birds make in trying to follow each other, and  the two
%direction-dependent sound speeds with
% and $\gamma$ and $v_2$ are additional phenomenological
%parameters with the dimensions of speed. Neither of them is generically equal to the
%mean speed $v_0$ of the flock. The speed $v_2$ will prove (as I'll show below) to be the ``magic'' speed that leads, if $v_s = v_2$, to faster wandering of the ``slow''
%bird; hence, the fact that $v_2\neq v_0$ is crucial to making the behavior of the``slow'' bird distinct from that of the other birds in the flock. In addition, where the 

Performing the integral over $\omega$ by
complex contour  techniques, and  using equation
(\ref{CLd=2}) for
$C_{\perp}(\vec{q}, \omega)$ in $d=2$, and (\ref{CTd=3}) for
$C_{\perp}(\vec{q}, \omega)$ in $d=3$, gives
\begin{widetext}
\begin{eqnarray}
C(\delta  t) = \left(\frac{v_s}{v_0}\right)^2{\Delta }\int {d^dq \over
(2\pi) ^d}   { \exp\left\{i\left((v_s-\gamma) q_{\parallel}
 \right)\delta t -\Gamma_\sigma (\vec{q}) \delta t\right\} 
\over \Gamma_\sigma(\vec{q})}
% \nonumber\\
\label{Ct1}
\end{eqnarray}
\end{widetext}
where  $\sigma = L$ in $d=2$, and  $\sigma = T$ in $d=3$.

Now clearly, if $v_s \neq \gamma$, the first term in the exponential is much larger, at small $q$, than the $\Gamma_L (\vec{q})$ term,  since the former scales as $q_{\parallel}$, while the latter scales as $q_{\parallel}^2$, in the dominant regime $q_{\parallel} \sim q_{\perp}^\zeta$. Hence, I can drop the latter term in the exponential, since it is small $q$'s which dominate the integral at large times (as we'll see in a moment).  Dropping this term,  and 
using the scaling law (\ref{Gamma scale}) for $\Gamma_{\sigma} (\vec{q})$, it is easy to see that rescaling the components $q_{\parallel}$ and $\vec{q}_{\perp}$ of $\vec{q}$ according to
\begin{eqnarray}
q_{\parallel} \equiv {Q_{\parallel} \over \delta t}, \quad \vec{q}_{\perp} \equiv {\vec{Q}\perp
\over \delta t^{1\over\zeta}}
\label{rescale1}
\end{eqnarray}
pulls all of the time dependence out of the integral $d^dq$, giving
the scaling law
\begin{eqnarray}
C(\delta t) \propto \delta t^{{1-d \over \zeta}-1+2}\propto \delta
t^{\phi(d)}
\label{Cscale1}
\end{eqnarray}
for $C(\delta t)$,
where I've defined
\begin{eqnarray}
\phi(d)\equiv{1-d \over \zeta}+1  ~~~.
\label{phidef}
\end{eqnarray}

It is important to note that this scaling law for $C(\delta t)$ only holds for $\delta t$ large, since it is only for such times that the hydrodynamic theory is valid. Thus, we are not concerned with any apparent divergences at short times 
that occur when 
(\ref{Cscale1}) is inserted into  (\ref{RW 3}). Divergences as  $t\rightarrow\infty$ {\it are} real, on the other hand; these will occur whenever $\phi(d)$, as given by equation (\ref{phidef}), is $>-1$. 

This happens in $d=2$, where equation (\ref{phidef}) gives $\phi(2)=-{2\over 3}$; inserting this into  (\ref{RW 3})
gives
\begin{eqnarray}
w^2 (t; v_s) \propto t^{{4\over3}}~~~~~~~(\rm{ordinary~ birds}, d=2),
\label{wordd=2}
\end{eqnarray}
a known  \cite{TT2} result  for
birds in  flocks of identical birds in $d=2$.
In $d=3$, on the other hand, equation (\ref{phidef}) gives $\phi(2)=-{3\over 2}<-1$. In this case, the integral over $t^{\prime\prime}$ in  (\ref{RW 3})
converges at large $|t^\prime -  t^{\prime\prime}|$, and so is dominated by short times, where, as just discussed, the hydrodynamic theory is invalid. Thus, the result of performing the integral over $t^{\prime\prime}$ in  (\ref{RW 3}) is a constant, independent of $t$ and $t^\prime$. As a result, the integral over $t^\prime$ in  (\ref{RW 3}) is just 
\begin{eqnarray}
w^2 (t; v_s) \propto t~~~~~~~(\rm{ordinary~ birds}, d=3),
\label{wordd=3}
\end{eqnarray}
another already known result\cite{TT2}. Note that these results apply not only to the identical birds that make up the bulk of the flock, but to {\it any} slow bird added to the flock whose  speed $v_s\ne\gamma$.

What happens if $v_s = \gamma$? Now things change dramatically,  since the $(v_s - \gamma)q_{\parallel} t$
term in the exponential in equation\ (\ref{Ct1}) vanishes.  As a result, 
I must keep
the
$\Gamma _\sigma$ term in the exponential in that equation.  Doing so, and
again using equation (\ref{Gamma scale}), I see that the rescaling to a new vector
variable  $\vec{Q}$ via
\begin{eqnarray}
q_{\parallel} \equiv {Q_{\parallel} \over \sqrt{\delta t}}
\label{rescale2a}
\end{eqnarray}
\begin{eqnarray}
\vec{q}_{\perp} \equiv {\vec{Q}_{\perp} \over \delta t^{1\over2\zeta}}
\label{rescale2b}
\end{eqnarray}
pulls all of the time dependence out of the integral when $v_s=\gamma$, and gives the scaling law
\begin{eqnarray}
C(\delta t)  = A   \delta t^{\phi(d)\over 2}  \quad ,
\label{Cscale2}
\end{eqnarray}
where $A$ is an unimportant constant.

Note that this decay is much slower than that given in equation(\ref{Cscale1}) for a bird
moving at a speed {\it other} than the magic speed $v_s = \gamma$.  Note
further that
since, in general, the mean speed $v_0$ of the flock differs from $\gamma$ (i.e.,
$v_0
\neq \gamma$) the identical birds that make up the bulk of the flock will exhibit
the more rapidly decaying correlations of equation(\ref{Cscale1}).  It is
only a bird
that differs from the generic members of the flock, specifically by flying at a
speed $v_s \neq v_0$, that has a chance to achieve $v_s = \gamma$, and the much
slower decay equation\ (\ref{Cscale2}).

This slower decay of velocity correlations leads immediately to faster lateral wandering for birds moving at this magic speed $\gamma$, 
 in both $d=2$ and $d=3$, since the exponent $\phi(d)\over 2$ is $>-1$ in both dimensions. As a result, the integral over $t^\prime$ and $t^{\prime\prime}$ is dominated by large times, and I find, by inserting (\ref{Cscale2}) into (\ref{RW 3}),
\begin{eqnarray}
w^2 (t ; v_s = \gamma) = A \int^t_0 dt^{\prime}\int^t_0
dt^{\prime\prime}\left|t^{\prime} -
t^{\prime\prime}\right|^{{\phi(d)\over 2}} \nonumber\\\propto t^{\frac{5}{2}-\left({1-d
\over2\zeta}\right)}= t^{{1\over\zeta}}~~~\rm{(slow ~bird, ~general}~ d<4) ,
\label{RW 7}
\end{eqnarray}
where in the last equality I've used equation (\ref{zetcan}) for $\zeta$.

Using equation (\ref{zetcan}) again, I get,  in $d=2$ spatial dimensions,
\begin{eqnarray}
w^2 (t ; v_s ) \propto
\left\{\begin{array}{ll}
t^{\frac{5}{3}},& v_s=\gamma~~,\\ t^{\frac{4}{3}},& v_s\ne\gamma~~(d=2),
\end{array}\right.
\label{wscaled=2}
\end{eqnarray}
while in $d=3$,
\begin{eqnarray}
w^2 (t ; v_s ) \propto
\left\{\begin{array}{ll}
t^{\frac{5}{4}},& v_s=\gamma~~,\\ t& v_s\ne\gamma~~(d=3),
\end{array}\right.
\label{wscaled=2}
\end{eqnarray}

In both dimensions, birds moving
at the magic speed $\gamma$ wander laterally much faster than birds moving at any
other speed. In particular, they wander much faster than the
``normal'' birds in
the flock.

What happens to a bird moving at a speed $v_s$ close to, but not equal to,
$\gamma$? For sufficiently short times, it should  be impossible to
tell that this bird was not moving at exactly the magic speed $\gamma$, and would
therefore expect to recover the $w^2 \propto t^{1\over\zeta}$ scaling I just
found for that case. For longer times, though, one would expect the difference
between the speed $v_s$ of the slow bird and the magic speed $\gamma$ to become
apparent, leading to a crossover to the $w^2 \propto t^{\frac{4}{3}}$ scaling
found in that case.

This proves to be precisely the case. To see this, note that the
rescaling equations
(\ref{rescale2a}), (\ref{rescale2b}) imply that the
integral in equation (\ref{Ct1}) for
$C(t;v_s)$ is dominated by parallel wavevectors
\begin{eqnarray}
q_{\parallel}^{dom} \propto t^{-\frac{1}{2}} \quad .
\label{qdom}
\end{eqnarray}
Comparing the $\delta v q_\parallel$ term in the exponent of equation (\ref{Ct1}) 
with
the
$\Gamma_L$ term there, where $\delta v \equiv v_s-\gamma$, and using the scaling law
(\ref{Dcross2}) for $\Gamma_L$ and (\ref{qdom}) for the dominant $q$,
I see that
$\delta v q_\parallel^{dom}\sim\Gamma_L(\vec{q}_{dom})$
at a time $t_c$ which scales according to $\delta v t_c^{-\frac{1}{2}}
\propto (t_c^{-\frac{1}{2}})^2$, which implies
\begin{eqnarray}
t_c(\delta v) \propto \delta v^{-2} \quad .
\label{crosst}
\end{eqnarray}
For smaller times ($t \ll t_c$), the $\Gamma_L$ term in equation
(\ref{Ct2}) dominates
the
$\delta v q_\parallel$ term, and  the $
w^2 (t)\propto  t^{1\over\zeta}$ law found in eqn (\ref{RW 7}) holds.  In the
opposite limit ($t\gg t_c$), the   $v_s\ne\gamma$ scaling  holds.

To summarize, for flocks  in $d=2$ spatial dimensions,
\begin{eqnarray}
w^2 (t ; v_s ) \propto
\left\{\begin{array}{ll}
t^{\frac{5}{3}},& t\ll
t_c(\delta v)~~,\\ t^{\frac{4}{3}},& t\gg
t_c(\delta v)~~,
\end{array}\right.
\label{wscaled=2}
\end{eqnarray}
while for $d=3$,
\begin{eqnarray}
w^2 (t ; v_s ) \propto
\left\{\begin{array}{ll}
t^{\frac{5}{4}},& t\ll
t_c(\delta v)~~,\\ t,& t\gg
t_c(\delta v)~~,
\end{array}\right.
\label{wscaled=2}
\end{eqnarray}
with $t_c(\delta v)$ given by (\ref{crosst}).

Numerical experiments could test this result by doing a series of simulations with added slow birds moving at different speeds. The magic speed $\gamma$ can be determined by fitting the above expressions. Alternatively, one can independently determine $\gamma$ by measuring the direction-dependent sound mode propagation speed in a simulation, and fitting the result to equation (\ref{cplusminus}), as was done in \cite{TT2}. 
An easier alternative for simulators who, understandably, do not wish to undertake this considerable numerical task, would be to simply simulate {\it exactly} the same model as was used in \cite{TT2}, with the addition of a single slow bird; in this case, one could simply use the value of $\gamma$ determined by \cite{TT2}.

In the (potentially simulate-able) case of spatial
dimensions $d\ge4$,  I find
simple ``drunkard's walk'' wandering ($w^2 (t ; v_s ) \propto t$)
independent of $v_s$.

I thank Sriram Ramaswamy, Yu-hai Tu, M. Cristina Marchetti, Hugues Chate, and  Francesco Ginelli for
many stimulatingly
combative discussions; Jean-Francois Joanny for a casual comment that clarified for me the correct way to treat the ordered phase of the flocking model; and the Max-Planck-Institut fur Physik komplexer Systeme, Dresden, and the Centre Emile Borel of the Institut
Henri Poincare, Paris, for their hospitality while a portion of this work was being done.  In
particular I thank Sylvie Lhermitte,	Claude Lucet, and Nitdavanh Sriratanakoul,
for getting me to the Paris airport
despite a transit
strike.


\begin{thebibliography}{000}


\bibitem{boids} C. Reynolds, {Computer Graphics} {\bf 21}, 25 (1987);
J.L Deneubourg and S. Goss, {Ethology, Ecology, Evolution} {\bf
1}, 295 (1989); A. Huth and C. Wissel, in {\em Biological Motion},
eds. W. Alt and E. Hoffmann (Springer Verlag, 1990)p. 577-590.
B. L. Partridge, {Scientific American}, 114-123(June 1982).

\bibitem{dictyo} W. Loomis, The Development of Dictyostelium
discoideum (Academic, New York, 1982); J.T. Bonner, The Cellular Slime Molds
(Princeton University Press, Princeton, NJ, 1967).

\bibitem{rappel1} W.J. Rappel, A. Nicol, A. Sarkissian, H. Levine,
W. F. Loomis, {Phys. Rev. Lett.}, {\bf 83}(6), 1247 (1999).

\bibitem {rappel2}H. Levine, W. J. Rappel, I. Cohen, {Phys. Rev.
E}, {\bf 63}, 17101 (2001).

\bibitem{Vicsek}
T.\ Vicsek, Phys.\ Rev.\ Lett. {\bf 75}, 1226 (1995); A.\
Czirok, H.\ E.\ Stanley, and T.\ Vicsek, J.\ Phys.\ A {\bf 30}, 1375
(1997); T. Vicsek, A. Czir\'{o}k, E. Ben-Jacob, I. Cohen, and O.  Shochet
Phys. Rev. Lett. {\bf 75}, 1226 (1995).





\bibitem{TT1} J.\ Toner and Y.\ Tu, Phys.\ Rev.\ Lett.\ {\bf 75},
4326 (1995).

\bibitem{TT2}Y.\ Tu, M.\ Ulm and  J.\  Toner, Phys.\ Rev.\ Lett.\  {\bf
80}, 4819 (1998).

\bibitem{TT3}J.\ Toner and Y.\  Tu, Phys.\ Rev.\ E {\bf 58}, 4828
(1998).

\bibitem{TT4}J.\ Toner,   Y.-h. Tu, and S. Ramaswamy, Ann.\ Phys.\
{\bf 318}, 170
(2005).

\bibitem{MW}
      N. D. Mermin and H. Wagner, Phys. Rev. Lett. {\bf 17}, 1133 (1966);
P. C. Hohenberg,
Phys. Rev. {\bf 158}, 383 (1967);  
\newline
N. D. Mermin, J. Math. Phys. {\bf
8}, 1061 (1967).

\bibitem{Chate1} See, e.g., G. Gregoire, H. Chate, Y. Tu, {Phys. Rev.
Lett.}, {\bf 86}, 556 (2001); G. Gregoire, H. Chate, Y. Tu, {Phys.
Rev. E}, {\bf 64}, 11902 (2001); G. Gregoire, H. Chate, Y. Tu,
{Physica D}, {\bf
181}, 157-171 (2003); G. Gregoire, H. Chate, {Phys. Rev. Lett.},
{\bf 92}(2), (2004).


\bibitem{Chate2}H. Chate, F. Ginelli, G. Gregoire, and F. Reynaud,
Phys. Rev. E {\bf 77}, 046113 (2008).

\bibitem{mom cons}S. Ramaswamy and R. A. Simha, Phys. Rev. Lett. {\bf 89}, 058101 (2002); Phys. Rev. A {\bf 306}, 262 (2002);  R. A. Simha, Ph. D. Thesis, Indian Institute of science, 2003; Y. Hatwalne, S. Ramaswamy, M. Rao, and R. A. Simha, Phys. Rev. Lett. {\bf 92}, 118101 (2004).

\bibitem{implicit} This should be thought of as an implicit
condition on
$v_0$, since
$\alpha$ and
$\beta$ can, in general, depend on $\left| \vec{v}\right|=v_0$.

\bibitem{Malt} J. Toner,  arXiv:1108.4392 (Condensed Matter), (2011).


\bibitem{kin theory} See, e.g., E. Bertin, M. Droz, and G. Gregoire, J. Phys. A: Math. Theor. {\bf 42}, 
445001 (2009) [arXiv:0907.4688];
E. Bertin, M. Droz, and G. Gregoire, Phys. Rev. E {\bf 74}, 022101 (2006) [cond-mat/0601038].

\bibitem{XY cubic}See, e.g., P. Calabrese and  P. Parruccini, Nucl.Phys. B {\bf 679}, 568 (2004).

\bibitem{slow}The speed $\gamma$  may,  in some  flocks, actually
be {\it greater} than the mean speed $v_0$ of the flock; hence, in some cases,
the ``slow'' bird may have to actually move {\it faster} than the
other birds. I will
describe the different bird in either case as the ``slow'' bird.

\bibitem{log}This ignores weak, logarithmic corrections in precisely four
dimensions; see \cite{TT1, TT2, TT3, TT4}; in practice, such
corrections should be
hard to see.


\bibitem{x sim 1} To see this, note that the argument of the scaling function $f_{\Gamma_L}$ is $O(1)$ in this regime; hence, $f_{\Gamma_L}$ itself should be roughly constant throughout this range of $\vec{q}$.

\bibitem{small_correction} Of course, this is not {\it exactly} true: the
slow bird, like the other birds, could make mistakes in following the local
direction of flock motion. However, such mistakes would simply add a
conventional, ``drunkard's walk'' lateral wandering (the RMS amplitude of
which would scale with time $t$ like $\sqrt{t}$) to the much larger 
wandering due to
following the flock that I am computing here.

%\bibitem{v2} What I call $v_2$ here was, confusingly, denoted $v_0$ 
%in references\cite{TT2,TT3,TT4}.


\end{thebibliography}
\end{document}